\documentclass{aa}  
\usepackage{graphicx}
\usepackage{txfonts}
\usepackage{natbib}
\newcommand{\starname}{LTT\,9779}
\newcommand{\planet}{LTT\,9779\,b}
\newcommand{\tess}{TESS}

\usepackage[breaklinks=true]{hyperref}
\hypersetup{
    colorlinks=true,
    linkcolor=blue,
    citecolor=blue,
    filecolor=magenta,      
    urlcolor=cyan
    }
    
\begin{document}

   %\title{A search for atomic elements in the atmosphere of LTT\,9779b using ESPRESSO}
   \title{A closer look at LTT\,9779b: The ESPRESSO endeavour to pierce the atmospheric veil}
   \authorrunning{R. Ramirez Reyes et al.}
   %\subtitle{I. Overviewing the $\kappa$-mechanism}

   \author{R. Ram\'irez Reyes,
          \inst{1}
          %\and
          James S. Jenkins\inst{2,3},
          Elyar Sedaghati\inst{4} ,
          J. V. Seidel\inst{4}\thanks{ESO Fellow}, 
          Yakiv Pavlenko\inst{5,6},
          E. Palle\inst{5,7},
          Mercedes L\'opez-Morales\inst{8},
          Douglas Alves\inst{1},
          Jos\'e Vines\inst{9},
          Pablo ~A. Pe{\~n}a R\inst{3},
          Mat\'ias R. D\'iaz\inst{10},
          Patricio Rojo\inst{1}
          %\fnmsep\thanks{Just to show the usage
          %of the elements in the author field}
          }
   \institute{Departamento de Astronom\'ia, Universidad de Chile,
              Camino el Observatorio 1515, Las Condes, Santiago, Chile\\
              \email{rramirez@das.uchile.cl}
         \and
         Instituto de Estudios Astrof\'isicos, Facultad de Ingenier\'ia y Ciencias, Universidad Diego Portales, Av. Ej\'ercito 441, Santiago, Chile
         \and 
         Centro de Astrof\'isica y Tecnolog\'ias Afines (CATA), Casilla 36-D, Santiago, Chile
         \and
         European Southern Observatory (ESO), Av. Alonso de Córdova 3107, 763 0355 Vitacura, Santiago, Chile
         \and
         Instituto de Astrof\'isica de Canarias (IAC), Calle V\'ia L\'actea s/n, E-38200 La Laguna, Tenerife, Spain
         \and
          Main Astronomical Observatory of the NAS of Ukraine, 27, Akademik Zabolotny Str., Kyiv, 03143, Ukraine
         \and
         Deptartamento de Astrofísica, Universidad de La Laguna (ULL), 38206 La Laguna, Tenerife, Spain
         \and
         Space Telescope Science Institute, 3700 San Martin Drive, Baltimore MD 21218, USA
         \and
         Instituto de Astronom\'ia, Universidad Cat\'olica del Norte, Angamos 0610, 1270709, Antofagasta, Chile
         \and
         Las Campanas Observatory, Carnegie Institution of Washington, Colina El Pino S/N, La Serena, Chile
             }

   %\date{Received September 15, 1996; accepted March 16, 1997}

  \abstract
  % context heading (optional)
  % {} leave it empty if necessary  
   {The proliferation of exoplanet discoveries, particularly within such exotic environments as the Neptune desert, challenges our understanding of planetary atmospheres undergoing intense irradiation. The unexpected discovery of \planet, an ultra-hot Neptune deep within this desert offers a prime opportunity for in-depth atmospheric studies. This research builds upon previous observations of LTT9779b from space-based telescopes, including the Transiting Exoplanet Survey Satellite (TESS),  Spitzer Space Telescope, and  CHaracterising ExOPlanet Satellite (CHEOPS), while incorporating new observations from the Very Large Telescope's (VLT) Echelle SPectrograph for Rocky Exoplanet and Stable Spectroscopic Observations (ESPRESSO) instrument to delve deeper into the atmospheric dynamics of this intriguing exoplanet. 
   Preliminary analyses suggest a metal-rich atmosphere alongside a notably high day-side geometric albedo that may imply the existence of silicate clouds. Furthermore, there appears to be minimal atmospheric escape, presenting intriguing contrasts to existing models of planetary evolution and atmospheric behaviour under extreme irradiation.
   }
  % aims heading (mandatory)
   {We aim to contribute to the broader understanding of atmospheric compositions and the mechanisms behind the survival of atmospheres in the Neptune desert through detailed spectroscopic analysis.  We  started by obtaining the transmission spectrum of LTT9779 b between 0.4 and 0.78 micrometres with ESPRESSO on the VLT.}
  % methods heading (mandatory)
   {Our analysis addressed systematics in ESPRESSO data across three distinct transit events, focusing on the sodium doublet and hydrogen alpha (H${\alpha}$). We also used the cross-correlation method with models that contain Na, K, FeH, TiO, and VO}
  % results heading (mandatory)
   {No statistically significant atmospheric signal was detected, with lower limits placed on the atmospheric metallicity  established at [Fe/H] $\geq 2.25$, which is $\geq 180\times$ solar. The non-detection is aligned with a high metallicity atmosphere scenario in a cloud-free model, suggesting a high mean molecular weight and a reduced atmospheric scale height.}
  % conclusions heading (optional), leave it empty if necessary 
  {We interpret the lack of any detection as evidence to support a very high metallicity for the planet's atmosphere. This would give rise to a high mean molecular weight and, hence, a low atmospheric scale height, rendering any signal too weak to be detected. Another possibility is the presence of high-altitude clouds or hazes that would suppress any signal from elements deeper in the atmosphere. These findings are consistent with recent consistent with recent James Webb Space Telescope (JWST) observations, which also report muted spectral features and suggest a high-metallicity atmosphere with clouds at high altitudes. Our results, together with those from JWST, support the hypothesis of a metal-rich atmosphere possibly obscured by clouds or hazes.}

   \keywords{planetary systems – planets and satellites: individual: \planet\ – planets and satellites: atmospheres – methods: observational – techniques: spectroscopic}

   \maketitle
%
%-------------------------------------------------------------------

\section{Introduction}

% Introduction general exoplanets, Neptune Desert. Introduction de LTT. (Indicate new papers in it)  
The discovery of large numbers of exoplanets, namely, planets orbiting stars other than the Sun, has opened up the possibility of carrying out some initial statistical studies on different planetary populations and their properties. One example is the discovery of the Neptune desert \citep{2011ApJ...727L..44S}, an area in the parameter space defined by a planetary radius and orbital period almost entirely devoid of Neptune-sized planets undergoing strong irradiation by their host star. This signature clearly cannot simply be explained  by observational bias \citep{2007A&A...461.1185L, 2013ApJ...763...12B}. However, the discovery of the first ultra-hot Neptune \planet\ deeply embedded in this area 
\citep{2020NatAs...4.1148J} challenges existing formation theories for the desert. It represents a unique laboratory for understanding not only the atmospheres of hot Neptunes but also what types of Neptune atmospheres  have managed to retain themselves in a high-irradiation environment. This planet was first observed to transit every 19 hours by the Transiting Exoplanet Survey Satellite 

(TESS; \citealt{2016SPIE.9904E..2BR}). It was confirmed from the ground by \citet{2020NatAs...4.1148J} using data from the High Accuracy Radial velocity Planet Searcher (HARPS) spectrograph \citep{2003Msngr.114...20M} and CORALIE 

\citep{2000A&A...359L..13Q, 2000A&A...356..590U}, the Swiss telescope's Cassegrain Fiber-Fed Echelle spectrograph. The measured mass and radius are $29.32 \pm 0.8$~M$_{\oplus}$ and $4.59 \pm 0.23$~R$_{\oplus}$, respectively, with additional parameters shown in Table \ref{table_star}. This places the exoplanet in the Neptune Desert region of the parameter space \citep{2011ApJ...727L..44S, 2018MNRAS.479.5012O}. Yet, \citet{2020NatAs...4.1148J} found that even after two billion years of evolution, the exoplanet still maintains an estimated atmospheric mass fraction of $\sim$9\%. From the same paper, the equilibrium temperature of \planet\  was determined to be $\sim$2000\ K (assuming zero Bond albedo and complete heat recirculation), indicating an atmosphere rich in gas-phase metals. 
Furthermore, the host star is of an absolute magniture of 9 in the V band, meaning that  detailed studies can be performed from the ground and space using current instrumentation.

%-------------------------------------------------------------
%                                             Simple A&A Table
%-------------------------------------------------------------
%
\begin{table}
\caption{LTT 9779 derived properties from  \cite{2020NatAs...4.1148J}}             % title of Table
\label{table_star}      % is used to refer this table in the text
\centering                          % used for centering table
\begin{tabular}{c c c c}        % centered columns (4 columns)
\hline\hline                 % inserts double horizontal lines
Parameter & Units & Value & Source \\    % table heading 
%\hline                        % inserts single horizontal line
%STAR & & & \\
\hline                        % inserts single horizontal line
   RA & J2000 & 23h54m40.60s & TESS \\      % inserting body of the table
   DEC & J2000 & -37d37m42.18s & TESS \\
   B & mag & $10.55 \pm 0.04$     & TYCHO \\
   V & mag & $9.10 \pm 0.02$    & TYCHO \\
   $T_{eff}$ & K &  $5445 \pm 84$   & SPECIES \\ 
   log $g$ & dex &  $4.43 \pm 0.31$   & SPECIES \\ 
   $[Fe/H]$    & dex &  $0.25 \pm 0.08$   & SPECIES \\
   $v$ sin $i$ & km $s^{-1}$ &  $1.06 \pm 0.37$   & SPECIES \\
   $M_{\bigstar}$ &  $M_{\odot}$ &  $1.03^{+0.03}_{-0.04} $   & SPECIES + MIST \\
   $R_{\bigstar}$ &  $R_{\odot}$ &  $0.95 \pm 0.01 $   & SPECIES + MIST \\
   $L_{\bigstar}$ &  $L_{\odot}$ &  $0.68 \pm 0.04 $   & YY + GAIA \\
   $Age$ &  Gyr &  $2.1^{+2.2}_{-1.4} $   & SPECIES + MIST \\
   
\hline                                   %inserts single line
\end{tabular}

\caption{\planet\ properties derived from \cite{2023ApJS..269...31E}}             % title of Table
\centering  
\begin{tabular}{c c c} 
\hline\hline                 
Parameter & Units & Value \\
\hline
 $a/R_{\bigstar}$ &  & $3.877^{+0.090}_{-0.091}$ \\
 $P$      &    days      & $0.79206410 \pm 0.00000014$  \\
 $M_{p}$  & $M_{\oplus}$ & $29.32^{+0.78}_{-0.81}$ \\
 $R_{p}$  &    $R_{J}$   & $0.421 \pm 0.021$ \\
 $e$      &              & $< 0.058$ \\
 $i$      &    deg       &  $76.39 \pm 0.43$ \\
 $T_{eq}$ & K & $1978 \pm 19$ \\
 \hline 
\end{tabular}
\end{table}

Observations from the Spitzer Space Telescope \citep{2004ApJS..154....1W} have been instrumental in uncovering the thermal dynamics and compositional hints of \planet's atmosphere through the detection of secondary eclipses and thermal phase curves across visible and infrared spectra, including phase curves from Spitzer and TESS \citep{2020ApJ...903L...7C, 2020ApJ...903L...6D}. Key findings include a dayside temperature of $1800 \pm 120$ K at $4.5 \mu$m  and a notably higher brightness temperature of $2305 \pm 141$ K at $3.6 \mu$m, alongside a nightside temperature of $700 \pm 430$ K. These measurements showcase significant thermal emission variability with wavelength, hinting at a complex thermal structure.

Further analyses revealed no thermal inversion, but did suggest the presence of strong molecular absorption, likely from CO, indicating a chemically rich atmosphere \citep{2020ApJ...903L...6D}. The absence of a significant offset in the planetary 'hotspot' and the phase curve's amplitude point to an inefficient heat redistribution \citep{2020ApJ...903L...7C}. This poses a challenge to  current general circulation models \citep{2011MNRAS.413.2380H, 2013A&A...558A..91P, 2019ApJ...883....4S, 2024ApJS..270...34C}.

The recent work by \citet{2023A&A...675A..81H}  identified an optical eclipse of \planet\ with a significant depth of $115 \pm 24$ppm using CHaracterising ExOPlanet Satellite (CHEOPS). These authors inferred an exceptionally high day-side geometric albedo of approximately 0.8. This high albedo, attributed to silicate clouds on the day side, necessitates an atmospheric metallicity of at least 400 times the solar value, given the extreme day-side temperatures observed \citep{2023A&A...675A..81H}. Such a metallicity level further corroborates the hypothesis of a metal-rich atmosphere. 

Collectively, these observations suggest a metal-rich atmosphere for \planet. This conclusion is particularly intriguing given its rarity in the Neptune desert, providing crucial insights into atmospheric composition and dynamics in extreme irradiation environments.

The scarcity of Neptune-sized planets in close orbits around their stars suggests significant evolutionary processes, including photoevaporation and migration mechanisms \citep{2018MNRAS.479.5012O, 2018MNRAS.476.5639I, 2022AJ....164..234V, 2024ApJ...962L..19V}. However, recent observations challenge these assumptions, as \citet{2023ApJS..269...31E} showed no clear evidence of atmospheric escape in \planet, suggesting a more complex atmospheric composition and dynamics. This leads to hypotheses that \planet \ might have started as a hydrogen-and-helium-rich planet that ended up with a higher mean molecular weight due to atmospheric escape processes.

The X-ray faintness of \planet's host star implies a less aggressive environment for atmospheric erosion, potentially aiding in the preservation of its atmosphere against photoevaporative forces \citep{2024MNRAS.527..911F}. This finding is puzzling given photoevaporation theories, but it could explain why \planet \ still retains its atmosphere even in such a highly irradiated environment. The relatively low high-energy irradiation from the star has not fully eroded the atmosphere.

Direct measurements or stronger constraints on \planet's atmospheric metallicity can help refine the formation and evolution models for this planet. \citet{2024ApJ...962L..20R} revealed muted spectral features, constraining the atmospheric metallicity to between 20 and 850 times the solar value. The presence of clouds at mbar pressures and the potential for silicate cloud condensation suggests complex atmospheric dynamics that might have contributed to retaining the atmosphere.

The metal-rich atmosphere of \planet \ is particularly intriguing given its rarity in the Neptune Desert, offering crucial insights into atmospheric composition and dynamics under extreme irradiation environments. Sodium and potassium detection, with their strong resonance lines in the visible spectrum, are key indicators of metal-rich atmospheres \citep{2001ApJ...553.1006B}. Previous detections in exoplanets such as HD 189733b, WASP-166 b, and WASP-127b have demonstrated the effectiveness of high-resolution spectrographs in identifying these signatures \citep{2008ApJ...673L..87R, 2015A&A...577A..62W, 2020A&A...644A.155A, 2022MNRAS.513L..15S}.

In this work, we describe the observation made with Echelle SPectrograph for Rocky Exoplanet and Stable Spectroscopic Observations (ESPRESSO) in Section 2. Section 3 describes the method to reduce the data. Section 4 describes the result in sodium and hydrogen-alpha (H${\alpha}$) lines, as well as the cross-correlation modelling. Finally, we discuss our findings in Section 5 and present our conclusions in Section 6.

%--------------------------------------------------------------------

\section{Observations}

The spectral time series of \planet\ used in this work was obtained from three separate transit events of the planet, with observations made on the nights of 3 November 2019, 10 November  2021, and 15 November  2021, all using %(Echelle SPectrograph for Rocky exoplanetS and Stellar Oscillations) 
the Echelle SPectrograph for Rocky exoplanetS and Stellar Oscillations %(ESPRESSO) spectrograph 
 
(ESPRESSO; \citealt{2021A&A...645A..96P}),
located at the 8m-class Very Large Telescope (VLT). Operating at a resolving power of $\mathcal{R}\sim$140\,000 in HR11 mode on UT3, and HR42 mode on UT2 and UT1, we were able to cover a wavelength range from 380\,nm to 788\,nm. The raw data were reduced with the standard European Southern Observatory (ESO) pipeline Data Reduction Software (DRS) v2.2.1 on ESOReflex 

The data reduction was conducted using the Reduction Software (DRS) v2.2.1 on ESOReflex \citep{2013A&A...559A..96F}. This pipeline performs several key processing steps, including bias and dark subtraction, echelle order localisation, flat-fielding, and wavelength calibration. Additionally, it assesses any cross-fibre contamination and the absolute efficiency of the instrument as a function of wavelength. The wavelength calibration utilises a Fabry-Perot etalon and a ThAr lamp. The pipeline then extracts the spectra into both 1D and 2D formats and computes the cross-correlation function (CCF) for each spectrum. This CCF is determined using a binary mask that closely matches the stellar spectral type, allowing for precise radial velocity (RV) measurements.
For our analysis, we utilised the 1D product (S1D\_SKYSUB\_A) generated by the ESO pipeline. The observations were part of programmes 2103.C-5063(A) (PI: Jenkins) and follow-up projects 108.22FQ.002 and 0108.C-0161(B) (PI: Ramirez), all of which are now public. While these observations share the same resolution, they differ in binning, with some systematic effects identified, particularly in 1x1 binning that was changed later (as discussed further in this paper). Detailed information about the observations is provided in Table \ref{table_obs}.

%-------------------------------------------------------------
%                                             Two column Table 
%-------------------------------------------------------------
%
\begin{table*}
\caption{ESPRESSO Observations log of LTT~9779}             
\label{table_obs}      
\centering          
\begin{tabular}{c c c c c c c c c}      
\hline\hline       
                     
Epoch \# & Date & Spectra & $T_{exp}[s] $& Airmass & Seeing [arcsec] & S/N@550nm  & Exec. Time & Readout mode \\ 
\hline                    
   Epoch 1 & 2019-11-03 & 54 (15 in, 37 out) & 148 &1.026-1.306 & 0.49-1.00 & 32-41 & 2:51:50 & HR 1x1\\
   Epoch 2 & 2021-11-10 & 57 (16 in, 38 out) & 140 &1.026-1.259 & 0.57-1.13 & 43-49 & 2:51:36 & HR 4x2\\
   Epoch 3 & 2021-11-15 & 56 (16 in, 44 out) & 140 &1.026-1.194 & 0.41-0.99 & 37-47 & 2:51:36 & HR 4x2\\
   
\hline                  
\end{tabular}
\end{table*}
%

%\subsection{Telluric Correction}

To correct the spectra for telluric absorption from atmospheric water and oxygen, we employed the ESO tool Molecfit (v. 3.0.3; \citealt{2015A&A...576A..77S}). This tool models Earth's transmission function using a line-by-line radiative transfer model, which fits the atmospheric profiles. These profiles are generated by combining a standard atmospheric profile with local meteorological data and dynamically retrieved altitude profiles for temperature and humidity at various pressure levels.

We first obtained a normalised synthetic stellar model from POLLUX (\citealt{2010A&A...516A..13P}), using a model that assumes a temperature of 5500 K, a surface gravity (log g) of 3.0, a mass of 1.0 $M_{\odot}$, a luminosity of $1.354 \log{L_\odot}$, and metallicity of 0.25 dex. This model is close enough to the real values to then be used to identify strong absorption lines in the stellar spectrum.

The next step involved shifting the spectrum to match the velocity reference of \starname\ by applying both radial and barycentric velocity corrections. We then masked regions with prominent stellar absorption lines where the line core drops below 96\% of the normalised flux. For the telluric correction, regions with strong telluric features were selected as input for Molecfit. This tool utilises the data Flexible Image Transport System (FITS) headers to retrieve information about the weather conditions on the specific night of observation. Notably, this includes regions around the Na doublet but excludes the core of the line.

\begin{figure*} 
   \centering
   \includegraphics[width=17cm]{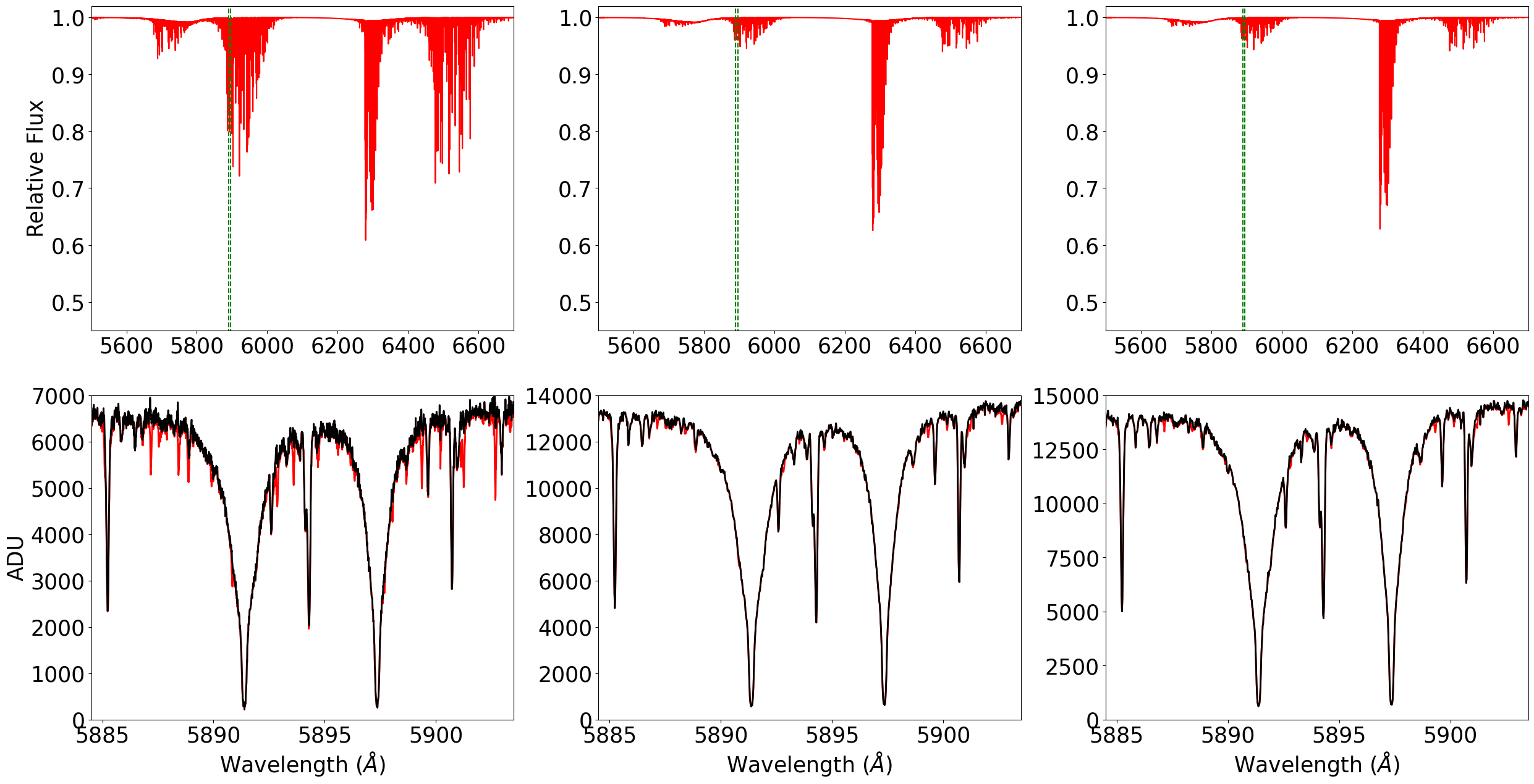}
      \caption{Example of the telluric correction of one exposure taken on each of the three nights. The red lines in the top panels are the Earth's transmission model from Molecfit for those exposures.  In the lower panels, the red curves are the uncorrected spectra, whereas the black curves are the corrected spectra. The panels from left to right correspond to nights 1, 2, and 3, respectively.}
         \label{Fig_Telluric_Correction}
   \end{figure*}

Finally, the observed spectra of \starname\ were divided by the model generated by Molecfit, effectively correcting the observed spectra for telluric contamination. This correction around the sodium doublet region is illustrated in Fig.\ref{Fig_Telluric_Correction}, which shows examples from each observed night.

Since the telluric model is a vital part of the reduction procedure, any anomaly could produce artificial signatures in the subsequent planetary transmission spectrum \citep{2015A&A...576A..77S, 2021MNRAS.502.4392L}. We also performed a visual inspection to verify the precision of the subtraction.

%-------------------------------------------------------------
%                 A Fig.as large as the width of the page
%-------------------------------------------------------------

\section{Method}

\subsection{Transmission spectra}

Before we can investigate the spectra for the transmission signal from the planet, additional effects such as the star's system RV and semi-amplitude, $K_{\star}$, the exoplanet's RV semi-amplitude, $K_p$, the Rossiter-McLaughlin effect (RM), \citep{1924ApJ....60...15R, 2012ApJ...757...18A}, and the centre-to-limb variation (CLV), \citep{2017A&A...603A..73Y} need to be addressed to assess the exoplanet's spectral signatures. 

The RM effect, which is manifested during the transit of a planet across a rotating star, results from the planet obscuring different segments of the stellar disc, thereby selectively blocking light from areas exhibiting distinct Doppler shifts. In our system, the star's rotational velocity was measured to be $1.06 \pm 0.5$ km/s \citep{2024MNRAS.527..911F}. Consequently, the anticipated RM effect is approximately $1.24$ m/s, a magnitude smaller than any atmospheric signals projected in our data. Thus, its influence on our measurements is expected to be minimal.

The CLV is an effect resulting from the variation in specific intensity as a function of distance from the centre of a star out to its limb. Our simulations indicate that the CLV effect is expected to produce a maximum signal amplitude of merely 0.005\% on the transmission signal, which falls well below the detectability threshold of our current methodology.

We note that while these effects have been quantified and are deemed negligible for this study, it is essential to consider them in other contexts where their impact might be more significant.
\
In the following sections, we focus on transmission spectra with no significant contamination from these effects.
We first corrected the telluric corrected spectra by the star's $K_{\star}$ to shift to the stellar rest frame. We shifted all the spectra to a common reference velocity using a simple linear interpolation. In this case, we selected the spectrum of observation number 30 (of 54) from the 2019 data and the first spectra from the data of 2021 as a common velocity reference for the rest of the spectra.

To perform the transmission spectroscopy, we used the star itself as a reference by co-adding all the spectra outside of the transit (where the planet was not crossing the stellar disc) to generate a master spectrum.

We calculated our first approximation to the transmission as follows:

\begin{align}
 \mathfrak{R}_i(\lambda) = \frac{f_i(\lambda)}{f_{\textrm{out}}}, 
\end{align}

\noindent where $f_i(\lambda)$ corresponds to each in-transit spectrum and $f_{out}$ is the master out of transit spectrum described above.

The next step is to consider the dispersion of every frame. Based on the approach to calculating a transmission spectrum in \citet{2015A&A...577A..62W}, we follow the implementation in \citet{2020MNRAS.494..363C}.%Therefore, following \citet{2020MNRAS.494..363C}, we define the following: 

\begin{align}
 \delta_i(\lambda) = 1 - \mathfrak{R}_i(\lambda, t),
\end{align}

where $\delta (\lambda)$ is the excess transit depth caused by the planet's atmosphere \citep{2020MNRAS.494..363C}. In this form, positive values will indicate that there is an extra atmospheric absorption present from the transiting planet. 

\ 
It is again important to mention that we are using the same star as a stellar reference. Therefore, any information in the absolute changes in the transmission spectrum in the continuum is lost.
The relation between $\delta$ and the scale height $H(\lambda)$ is as follows:

\begin{align}
\delta(\lambda) = 1 - \mathfrak{R}(\lambda) \simeq \frac{2 R_p H(\lambda)}{R^2_{\ast}}. 
\end{align}

%-------------------------------------------------------------
%                 A Fig.as large as the width of the column
%-------------------------------------------------------------
   \begin{figure*}
   \centering
   \includegraphics[width=17cm]{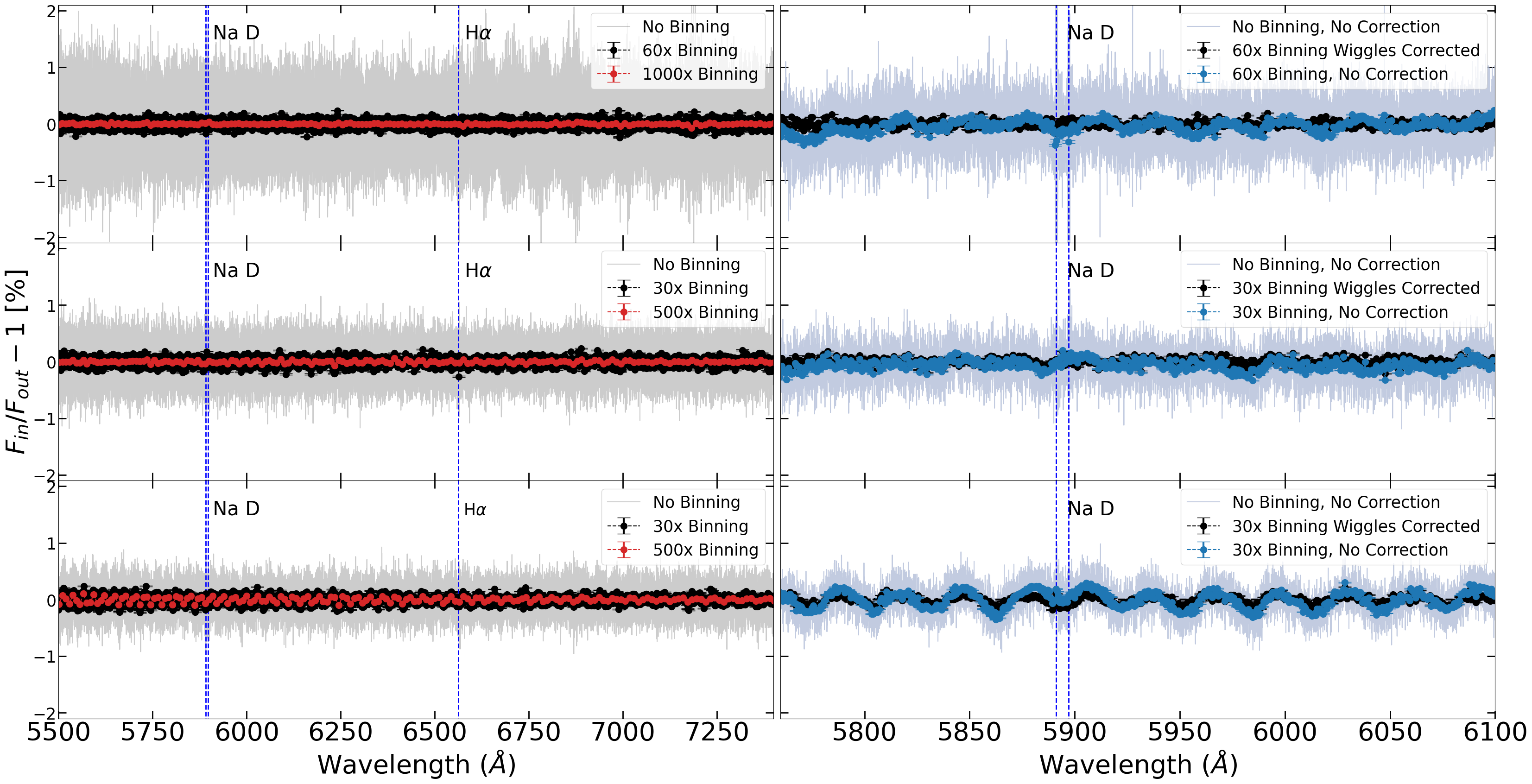}
      \caption{Wiggle-corrected transmission spectra of LTT~9779b across the full wavelength range. The panels present transmission spectra from three epochs, obtained with the red arm of ESPRESSO. The grey line shows the raw data, the black lines represent the same data binned by a factor of 30, and the orange dots correspond to binning by a factor of 500. The sodium doublet and H$\alpha$ features are marked by blue vertical lines for reference. The top panel displays the 2019 data, while the middle and bottom panels correspond to the 2021 observations. The right panels follow a similar format, comparing spectra with and without the low-frequency wiggle correction.}
      \label{Overall_Transmission}
   \end{figure*}

Although the spectra have been processed with the standard ESO pipeline, where the sky is subtracted to avoid possible Na contamination, and finally cleaned off the telluric lines using Molecfit, subtle differences remain in the expected continuum.

In their studies, \citet{2020A&A...644A.155A} and 
\citet{2021A&A...647A..26C} identified sinusoidal spectral distortions, termed 'wiggles,' which are instrumental in origin. These features, including variations with an amplitude of about 0.3\% and finer oscillations of 0.1\%, are primarily attributed to internal reflections within the P4/P5 fixed prism of the ESPRESSO instrument. Such effects, observed in particular with the UT3 telescope, are critical for accurate spectral analysis and have been addressed using specific correction methodologies in subsequent observations. For a detailed discussion of the wiggles' characteristics and their mitigation, we refer to the referenced works.
To correct these effects, we proceeded with a two-step approach. 

First, we removed the long-period oscillation (wiggles) present in the transmission spectra using a Savitzky–Golay filter, as implemented in the SciPy library \citep{2020NatMe..17..261V}. Additionally, we tested a Gaussian process (GP) to remove the high-frequency oscillation (mini-wiggles). The overall result is shown in Fig.\ref{Overall_Transmission}, where the transmission spectra from epochs 1 to 3 are shown from top to bottom. The specific configuration of this reduction is described below.

We begin by binning the transmission spectrum data by a factor of 100 to decrease the white noise and reveal the overall oscillation. On this binned data, we applied the Savitzky-Golay filter, which has the advantage of preserving the original shape and features of the signal better than other types of filtering approaches, such as methods based on a moving average  \citep{Guinon2007MovingMathcad}. 
We meticulously optimised the window length of the filter to minimise residual oscillations, ultimately setting it at 2000 data points. Additionally, the filter order was established at three. Despite the data not being equally spaced, we observed that methods accounting for these discrepancies yielded comparable results.

The refined function derived from this process was then used to mitigate oscillations in the unbinned data through straightforward subtraction. Parallel tests were conducted with RASSINE \citep{2020A&A...640A..42C}, a sophisticated tool designed for normalising stellar spectra. RASSINE's utility in our context lies in its implementation of the Savitzky-Golay filter for the removal of wiggles or oscillatory features in spectral data, similarly to our approach. %These tests with RASSINE corroborated our findings, underscoring the effectiveness of the Savitzky-Golay filter in our analytical methodology. 

% Pato: I figure will be nice to compare RASSINE with Savitsky-Golay

The correction was made with a simple subtraction in the Savitzky-Golay corrected residual. The SG correction was done first to avoid an overfitting. Afterwards, we modelled the spectra after applying the SG filter with a GP model that uses a quasi-periodic oscillation kernel constructed using simple harmonic oscillators. To proceed, the dataset was binned by a factor of 50. We first tested with the George package \citep{2015ITPAM..38..252A} and a Matern 32 kernel. We transitioned to using Celerite \citep{2017AJ....154..220F} due to its superior convergence speed compared to George for our specific objectives. Additionally, we adopted quasi-periodic oscillators, which exhibit enhanced performance in extrapolating across gaps in the data. This approach was instrumental in preventing any perturbation of the sodium line cores.

Another important factor is a physical one, as the planet moves around the star.
From the ingress to the egress, its velocity changes from -30 to 30 km/s, corresponding to $\sim$1\r{A}, or $\sim$120 pixels. Therefore, we applied that series of shifts to the planet's spectra in order to place all residual spectra in the planetary rest frame.  The combination of the frames taken in transit was weighted by its dispersion, similar to the transit master spectrum procedure explained above.  The combination between epochs was carried out with a simple mean.

\subsection{Cross-correlation}

Using the transmission spectroscopy technique, we searched for Na, K, and H${\alpha}$ in each of the corrected transmission spectra for each night. Other lines from V, Fe I, Ca or CO are expected to be weaker in the spectrum and therefore harder to detect, requiring another method to boost the sensitivity. The cross-correlation between the spectra and a planetary spectral model has previously been used to detect these elements, further boosted by stacking the signal from all the absorption lines \citep{2010Natur.465.1049S, 2018Natur.560..453H, 2019A&A...627A.165H}.
\ 

The cross-correlation technique requires a model of a theoretical transmission spectrum of one or multiple elements. We used the publicly available code petitRADTRANS \citep{2019A&A...627A..67M, 2020A&A...640A.131M} to obtain atmospheric models with a variety of elemental and molecular abundances. We refer to Figure \ref{Fig_MassFraction} for the list of elements and abundances. This code uses the chemical equilibrium abundance as a first approximation, which is appropriate for a hot planet such as this one.

Subsequently, we convolved the template with a narrow Gaussian filter, sourced from SciPy \citep{2020NatMe..17..261V}, with an FWHM of 4.5 pixels (0.045Å), aligning with the line spread function of ESPRESSO. Post-convolution, we normalised the spectra. This normalisation involved separating the data into distinct wavelength and flux bins, selecting only the higher flux values from these bins to establish a simple polynomial function for normalisation.
\ 
Following \citet{2020MNRAS.494..363C}, we defined the CCF as:

\begin{align}
 g_{\rm CCF}(\lambda, t) = \frac{\sum_i m_i(\lambda)w_i(\lambda,t)R_i(\lambda,t)}{\sum_i m_i(\lambda)w_i(\lambda,t)},
\end{align}
 
\noindent where $m_i$ is the model from petitRADTRANS, $w_i$ is the weight assigned to each bin, and $R_i$ is the transmission defined previously. The weights were built according to the stability of that wavelength; that is,  $ 1/ \sigma $, where $ \sigma$ is the geometrical addition of the signal time variation, previous error propagation, and the normalised magnitude of the telluric line correction, while $ R_i $ is the residual vector that is correlated with the expected absorption features of the exoplanet's atmosphere. We refer to signal time variation as the standard deviation measured through the frames over time for every dataset at a different wavelength. The error propagation is the photometric error from the ESO pipeline, and the magnitude of the telluric lines refers to the fact that the flux value is increased because of the atmospheric absorption correction. Therefore, the error considers the observed flux, the derived error from the Molecfit model, and the subsequent division.  
%We perform the cross-correlation within a velocity span of -400 to 400 km/s in steps of $\sim 1.0$ km/s.
We performed the cross-correlation within a velocity span of -400 to 400 km/s in steps of 1.0 km/s. Initially, we used a step size of 0.2 km/s, which resulted in oversampling for ESPRESSO’s natural pixel size of 0.5 km/s. Following analysis with step sizes of 0.5, 1.0, 2.0, and 4.0 km/s, we found no significant differences in the results across 0.2, 0.5, and 1.0 km/s. Therefore, we opted for a step size of 1.0 km/s to balance resolution and computational efficiency.

\

To analyse the transmission spectrum, we cross-correlated each frame and looked for the expected 'moving trail' indicative of planetary detection. We enhanced the signal-to-noise ratio (S/N) by co-adding all frames in the planet's velocity reference frame. We applied a 2D median filter using the medfilt2d function from SciPy with a kernel size of 51 pixels to remove any apparent systematic slopes unrelated to the transit. The magnitude of this correction was smaller than the dispersions present across all data sets.

\subsection{Detection limit calculations}

%-------------------------------------------------------------
%                 A Fig.as large as the width of the column
%-------------------------------------------------------------
   \begin{figure*}
   \centering
   \includegraphics[width=17cm]{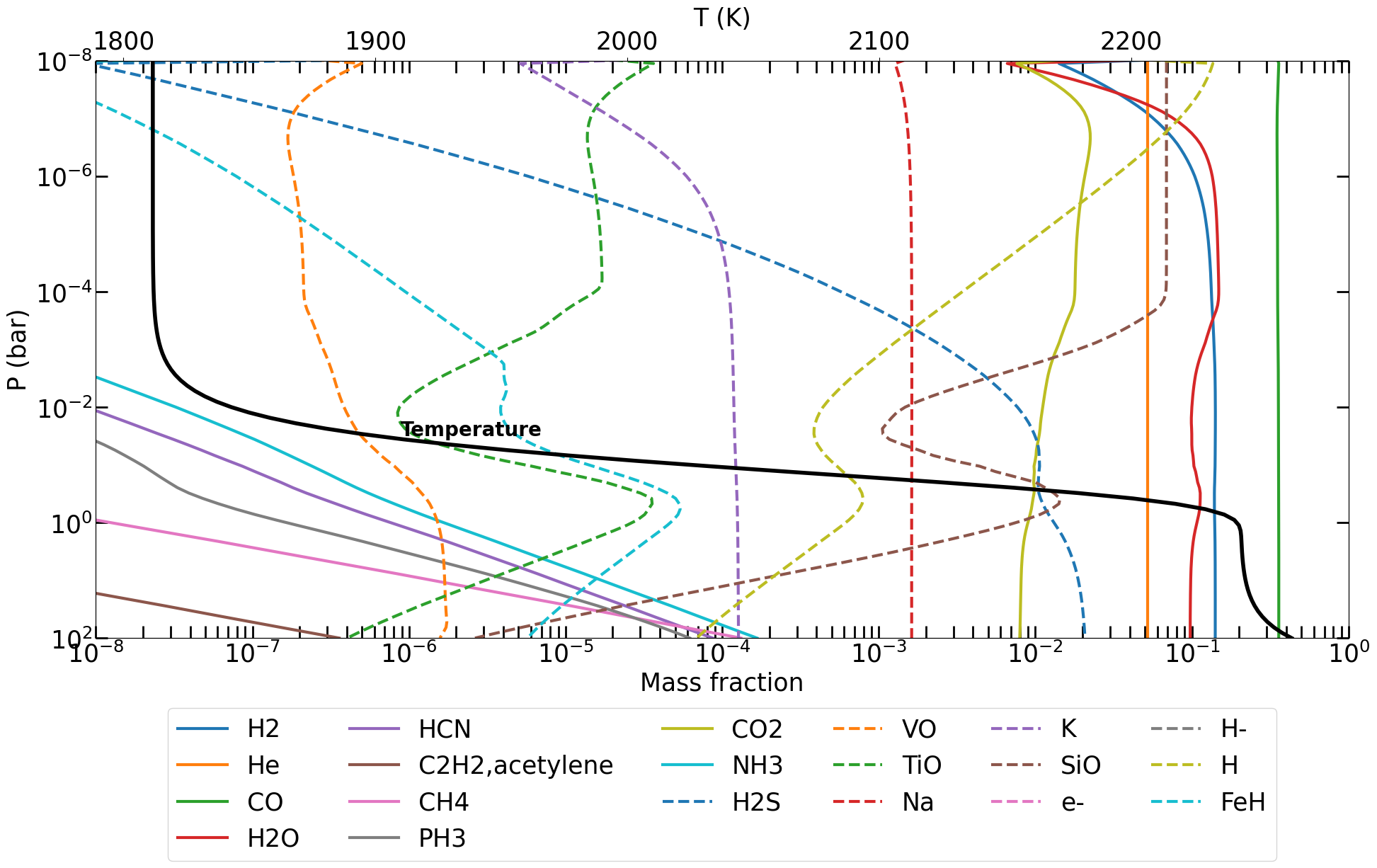} 
      \caption{Calculated mass fractions of various atomic elements in an exoplanetary atmosphere similar to that of \planet, and considering different planetary pressures. The calculations are based on a specified metallicity with [Fe/H] = 2.5. Notably, the atmospheric temperature made with guillot model from petitRADTRANS (Version 2) is delineated by a black line and a separate scale located at the top of the plot for easy reference. }
         \label{Fig_MassFraction}
   \end{figure*}

To determine our detection limits, we performed several tests that allowed us to define the upper limits of these so-called 'non-detections'. We used petitRADTRANS to calculate an expected signal with different atmospheric planetary abundances.
\

We assumed that the planet's atmosphere is predominantly composed of $H_2$ and He, which contribute to Rayleigh scattering and continuum opacities. Additionally, for all the species included in the models, shown in Fig.\ref{Fig_MassFraction}. The petitRADTRANS code is employed for line-by-line radiative transfer calculations in a plane-parallel atmosphere, assuming local thermal equilibrium (LTE).

Our model adopts a 1D temperature profile, with the pressure-temperature (P-T) profile and elemental abundances defined at various atmospheric layers or radii within the planet, assuming equilibrium chemistry for all the species we are including. For atmospheric stratification, we utilized 130 layers, distributed evenly on a logarithmic scale, extending from $10^2$ bar to $10^{-10}$ bar. The Guillot temperature model, as implemented in petitRADTRANS \citep{2010A&A...520A..27G}, was applied with parameters, $T_{\mathrm{int}} = 200$ K, $T_{\mathrm{equ}} = 2000$ K, $\gamma = 0.4$, and $\kappa_{\mathrm{IR}} = 0.01 \, \mathrm{cm}^2\, \mathrm{g}^{-1}$. The mean molecular weight (MMW) at each pressure layer is computed based on the respective mass fractions. The model's wavelength coverage spans from 0.4 to 0.8 $\mu$m.

In the final step, the model output is convolved to account for planetary rotation, the line spread function to align with the resolving power of ESPRESSO ($R \approx 140,000$), and smearing caused by the dynamical movement of the system throughout the observations. This convolution employs a Gaussian function for rotational broadening and a boxcar function for time integration.

To determine the upper limit for detection in our analysis, synthetic data were generated using parameters analogous to those previously described but with variation in metallicities, ranging from -1 to +3 in the [Fe/H] scale, with $\alpha$/Fe constant. Consequently, not all models in this set would end up dominated by $H_2$ and He. This tailored model was then injected into a synthetic flux, derived from the average stellar spectrum observed outside of the transit. Additionally, photon noise was incorporated into this synthetic flux to align the  S/N with that measured in the in-transit flux.

%-------------------------------------------------------------
%                 A Fig.as large as the width of the column
%-------------------------------------------------------------
   \begin{figure}
   \centering
   \includegraphics[width=\hsize]{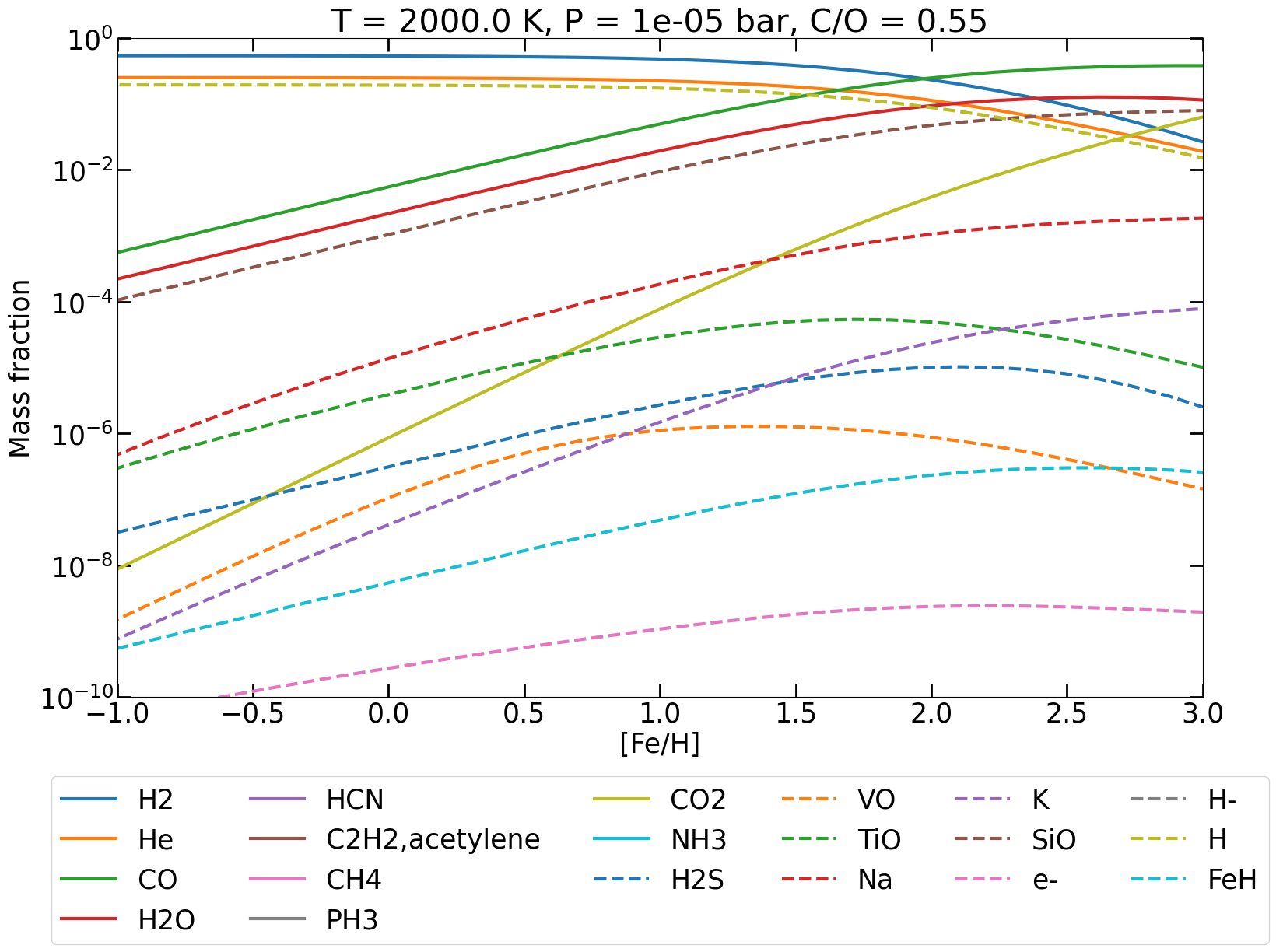} 
      \caption{Calculated mass fractions of various atomic elements in an exoplanetary atmosphere under differing metallicity conditions, utilising PetitRADTRANS. These fractions were determined at a temperature of 2000K and a pressure of 1 millibar. The model assumes a carbon-to-oxygen (C/O) ratio of 0.55.}
         \label{Fig_MassFraction2}
   \end{figure}

\section{Results}

\subsection{Transmission spectrum of \planet}

\begin{figure}
   \centering
   \includegraphics[width=\hsize]{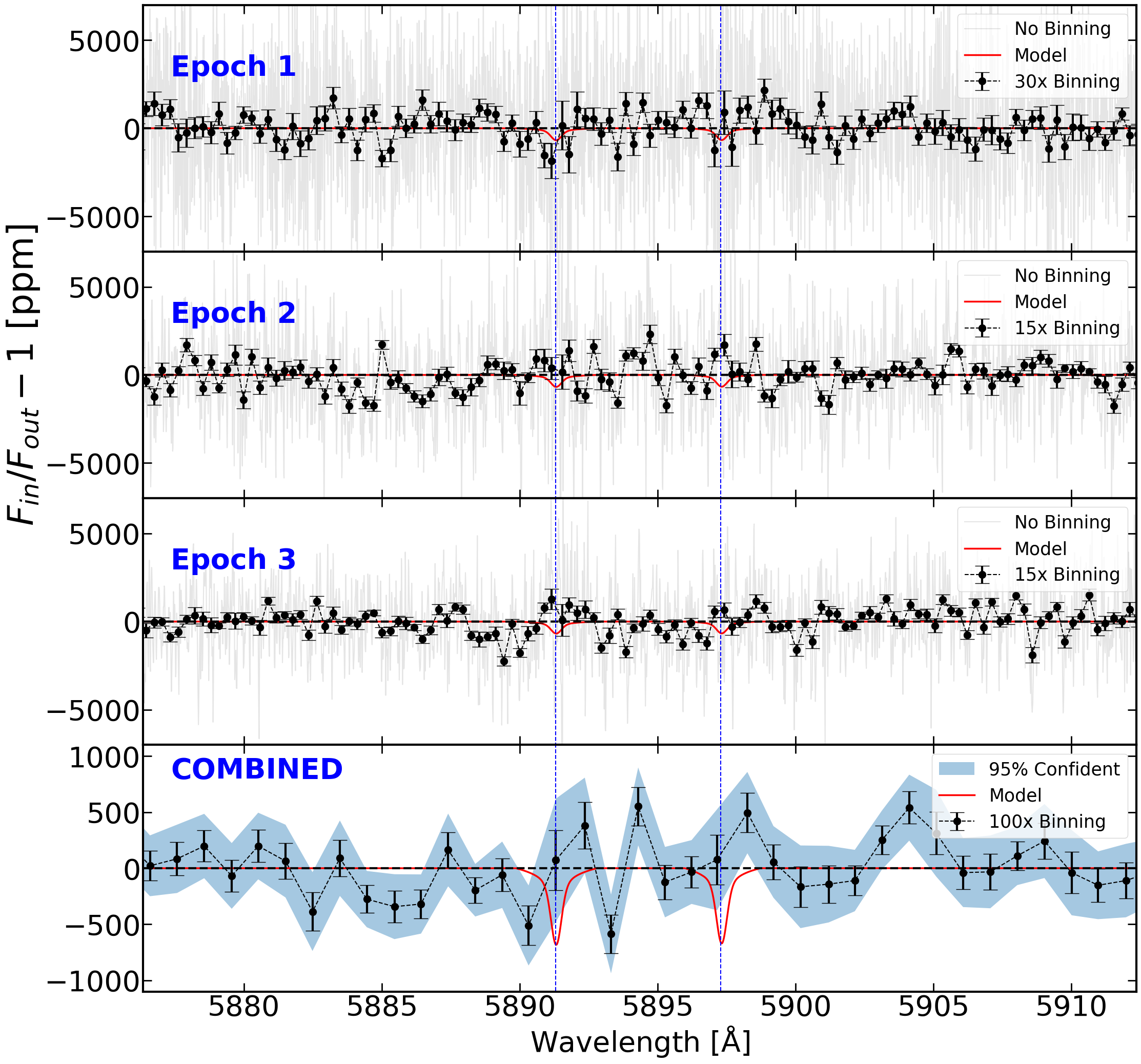}
      \caption{Transmission spectra around the sodium doublet for the three epochs. The three upper panels are the transmission spectrum of each individual epoch.  Grey lines are the transmission spectrum. Black dots are the data binned by a factor of 30 and 15, for the upper and middle two panels respectively, error bars are calculated as the standard deviation of the data inside the bins, and red lines are the predicted signal of the petitRADTRANS model. The bottom panel shows the combined data at a different binning. The light blue area shows the 95\% confidence region assuming a Student’s t-distribution. Vertical blue lines indicate the core of the sodium doublet in reference to the host star. 
}
         \label{Fig_Sodium_NoDetection}
   \end{figure}

We did not detect any statistically significant absorption from the atmosphere of \planet. We performed the transmission analysis described above to search for excess in-transit absorption of Na, K, and H$\alpha$ lines, however no signal was present. Fig.\ref{Fig_Sodium_NoDetection} shows the resulting transmission spectra around the sodium doublet. The bottom panel shows the co-added spectrum with the highest S/N ratio. The standard deviation of the binned transmission spectrum in the continuum, from $5875 \r{A}$ to $5888 \r{A}$ and from $5897 \r{A}$ and $5915 \r{A}$, is 120 ppm. The signal variation detected around the area of the D1 $\lambda_0 = 5889.95 \r{A}$ and D2  $\lambda_0 = 5895.92 \r{A}$ lines is consistent with the increased noise expected in those areas due to the flux decrease.  The single data point that is slightly raised at the D2 line core is consistent with that deviation; in fact, the data point disappeared as the binning setup changed.

%-------------------------------------------------------------
%                 A Fig.as large as the width of the column
%-------------------------------------------------------------
%

%-------------------------------------------------------------
%                 A Fig.as large as the width of the column
%-------------------------------------------------------------
   \begin{figure}
   \centering
   \includegraphics[width=\hsize]{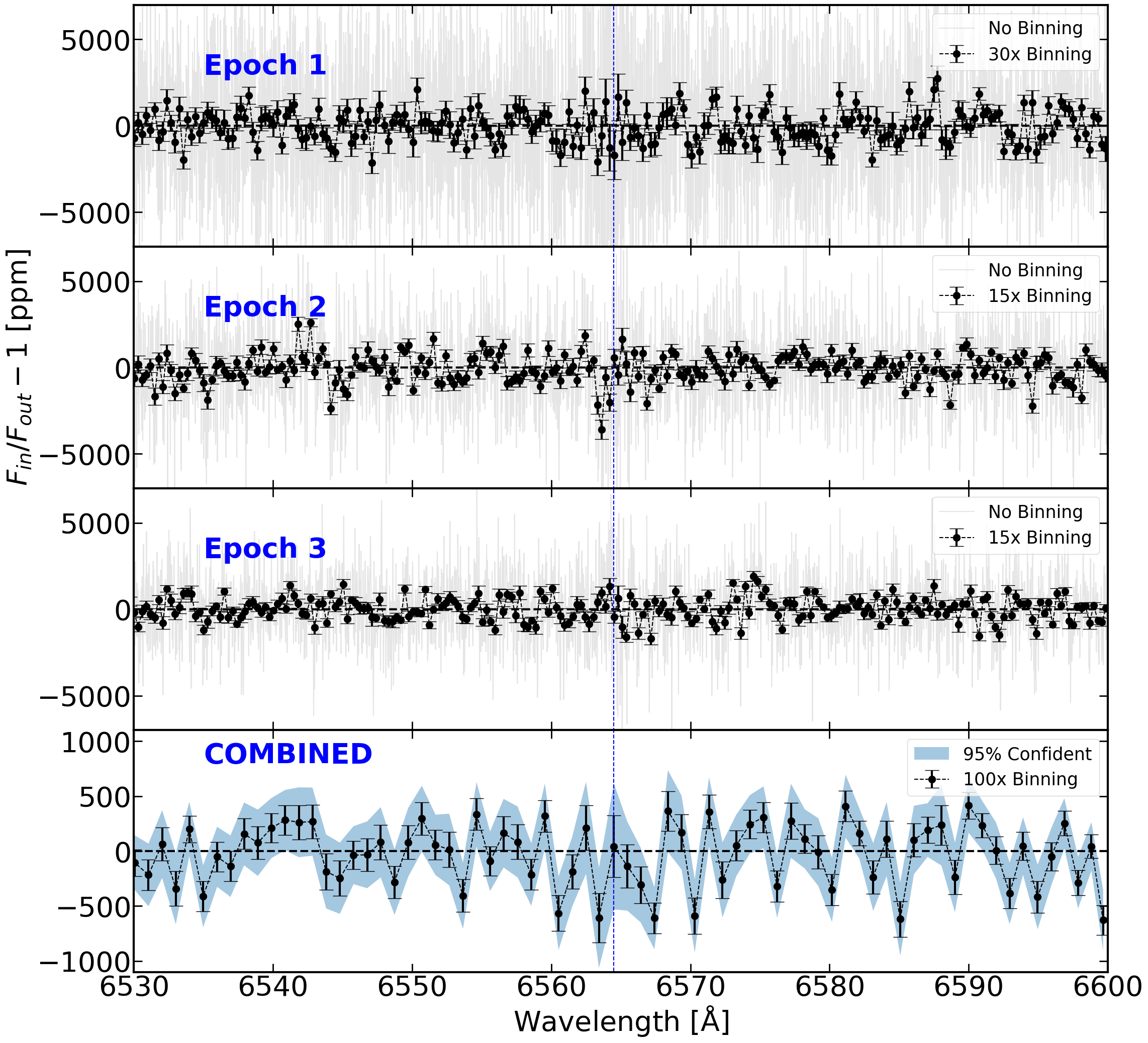}
      \caption{Transmission spectrum around the H${\alpha}$ line. The upper panel shows the transmission spectrum, with the grey line the complete data, and the black line the same data binned by a factor of 30. Error bars were calculated as the standard deviation of the bin. The lower panel shows the original spectrum, where we can identify the strong H${\alpha}$ absorption feature.}
         \label{Fig_H_NoDetection}
   \end{figure}

In the context of the H${\alpha}$ line, as depicted in Fig.\ref{Fig_H_NoDetection}, the standard deviation in the continuum near this line is 210 ppm, specifically in the ranges of $6550 \r{A}$ to $6560 \r{A}$ and $6568 \r{A}$ to $6578 \r{A}$. Consequently, at $\lambda_0 = 6563 \r{A}$, we observed no significant deviation from this standard, indicating the absence of a detectable signal in the core of the H${\alpha}$ line. This result implies a lack of clear evidence for photoevaporation. As a reference, a simple estimation of the scale height at 2000K and a mean molecular weight of 2.33 is $\sim 500 km$ or $\sim 45 ppm$. An escaping atmosphere is expected to significantly extend beyond the scale height, potentially reaching the Roche lobe, as observed in such cases as HD 209458b where the atmosphere expands to several planetary radii \citep{2003ApJ...598L.121L}. In this context, a simple estimation of the Roche lobe transit depth is approximately 6000 ppm (0.6\%), although this value should be treated as a rough approximation. Regarding the potassium line, challenges arose due to the proximity of telluric lines interfering with the spectral data. Combined with a low S/N, these factors hindered the extraction of a reliable signal in this spectral region.

The expected transmission spectrum signal modelled using petitRADTRANS with an equivalent planet atmospheric metallicity of [Fe/H] = +1.0~dex, a temperature of 2000 K, a mean molecular weight of 2.33 g/mol, and a C/O ratio of 0.5 shows a predicted signal of $\sim 500$ ppm at the core of the Na lines (red line in Fig.\ref{Fig_Sodium_NoDetection}). This signal is lower than the confidence level calculated at the line core (light blue area in the bottom panel of Fig. \ref{Fig_Sodium_NoDetection}). However, the confidence level at the line core is also affected by correlated noise, as observed, meaning that even though the light blue area suggests a confidence level approaching three sigma, this is insufficient to make a strong claim. Therefore, despite the data showing no transmission observed in the sodium doublet, it does not rule out the presence of a sodium signal with those properties. We we discuss this result further in our analysis in Section 5, .
\ 

In the case of H${\alpha}$, the signature of an evaporating atmosphere may have been expected, as it could be undergoing a Roche-Lobe overflow \citep{2017ApJ...835..145J} process and photoevaporation \citep{2020NatAs...4.1148J}. In that case, a stronger absorption signal than that predicted for sodium would be expected, which is not present in our results.

\subsection{Cross-correlation with atomic species}

\begin{figure}
   \centering
   \includegraphics[width=\hsize]{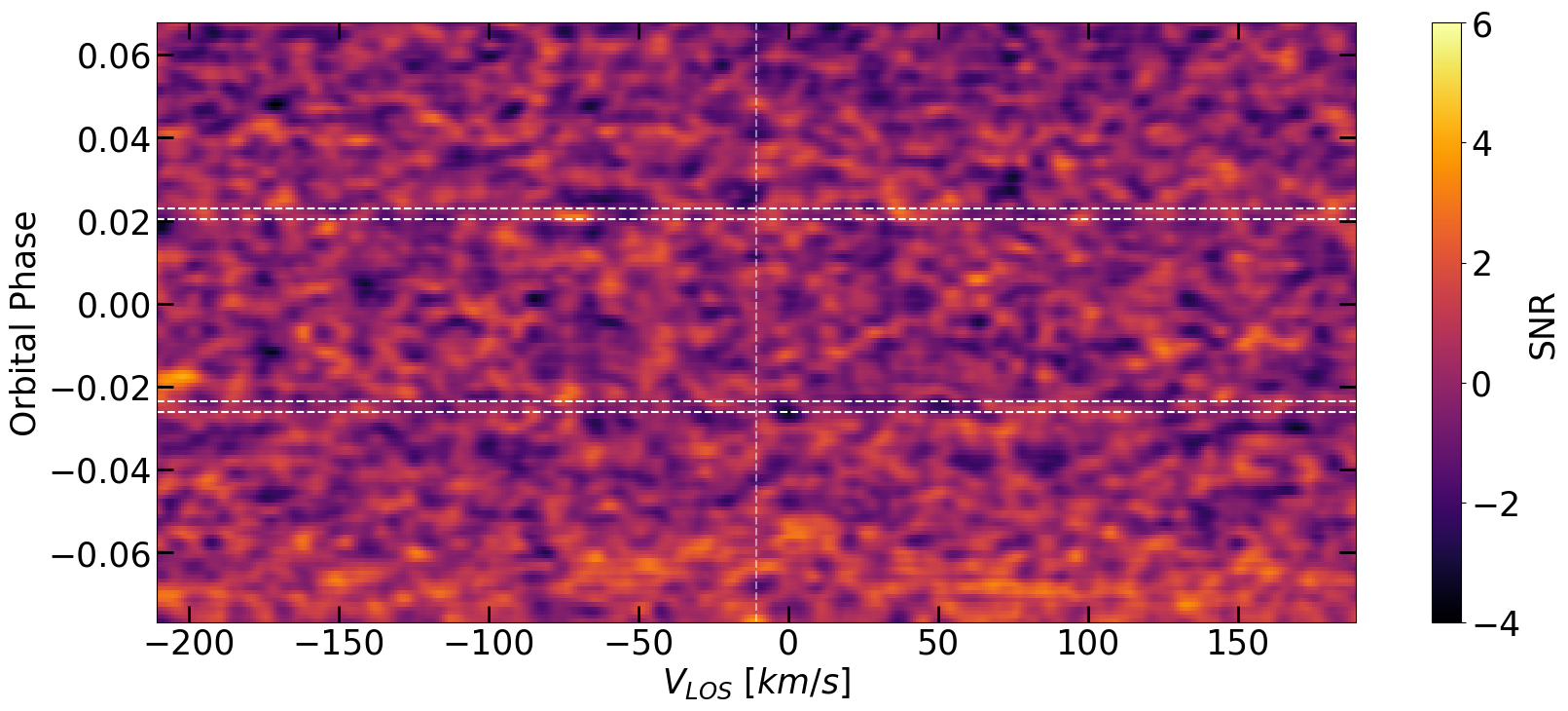}
      \caption{Cross-correlation signal made with a model that contains the signature of Na, K, FeH, V, and TiO with the combination of the transmission spectrum of the three epochs. The white dotted line indicates the moment of ingress and egress and when the full transit starts and ends. The velocity reference of the star is marked as a vertical line at -10.63 km/s. The x-axis indicates the offset relative to the line of sight velocity of the stellar system ($V_{LOS}$)}
         \label{CC1}
   \end{figure}

In our analysis of the CCF, we did not observe detectable signatures of VO, TiO, or FeH (similarly to the case with Na and K). A comprehensive CCF model that encompassed all these elements, including Na and K, was applied; however, it yielded no significant detections. The cross-correlation analysis was conducted on every frame, covering both in-transit and out-of-transit periods. Figure \ref{CC1} illustrates the S/N levels, taking into account the noise characteristics of the CCF during out-of-transit phases. It is important to note that even though the assumption of  Gaussian noise might not be an exact representation, it serves as a useful approximation to evaluate the strength of the signal. This approach indicates that the dispersions measured in the CCFs both outside and inside the transit are comparable, leading to the conclusion that there is no evident detection of the aforementioned elements.

%-------------------------------------------------------------
%                 A Fig.as large as the width of the column
%-------------------------------------------------------------

\begin{figure}
   \centering
   \includegraphics[width=\hsize]{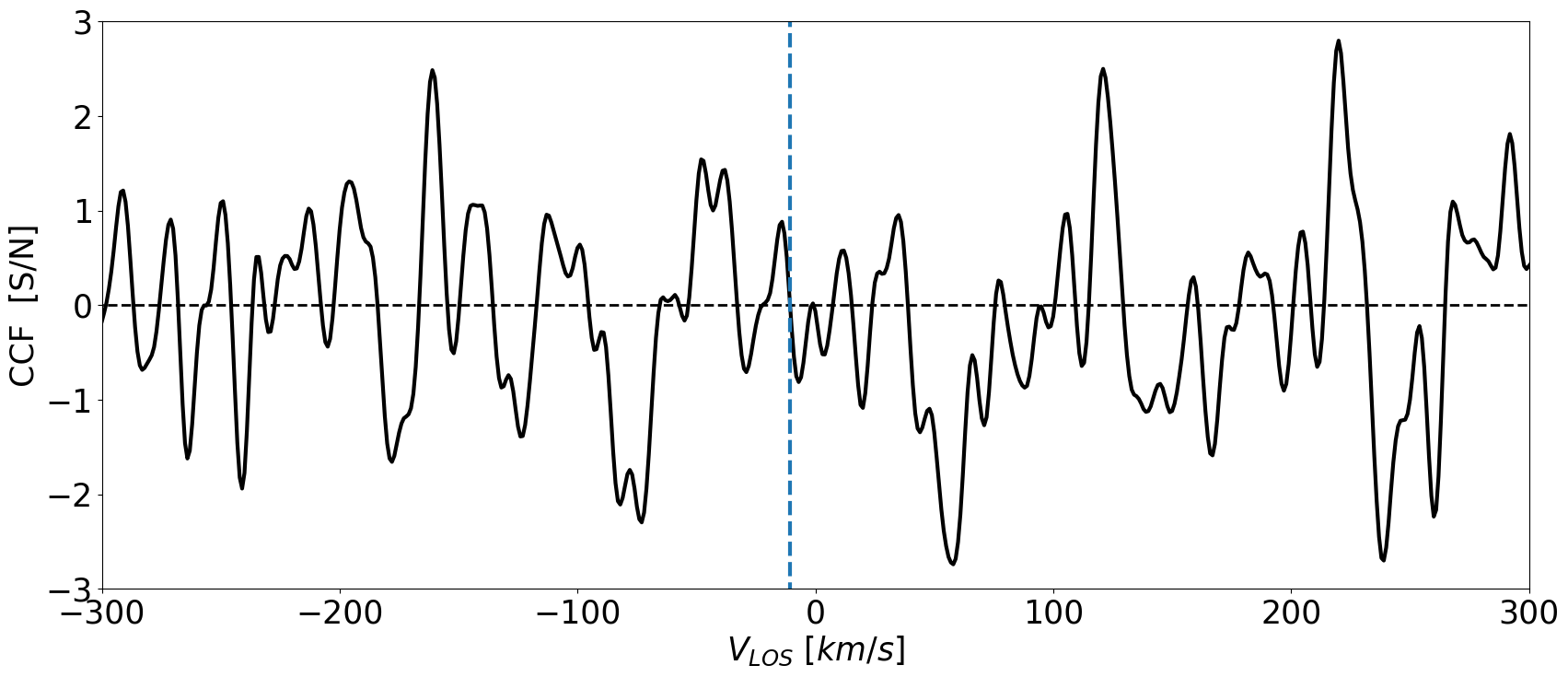}
      \caption{Cross-correlation signal made from a model that includes Na, K, Fe, V, and TiO with the combined spectrum from the three epochs. Frames were combined considering the Doppler shift of the planet. Positive values represent a positive correlation with the model, whereas negative values are anti-correlated. The units are in given as S/N with respect to the overall signal. The velocity reference of the star is marked as a blue line at -10.63 km/s. The x-axis indicates the offset relative to the line of sight velocity of the stellar system ($V_{LOS}$)}
         \label{CCF2}
   \end{figure}

We can collapse all CCF frames inside transit to one single CCF, thus increasing the S/N. To accomplish this, the changes in the planet's velocity (or Doppler shift) must also be considered to combine the frames. This is particularly important for planets so close to the host star as LTT 9779b. In this case, the maximum Doppler shift is $\sim$ -30 km/s at ingress and $\sim$ 28 km/s reaching the egress. Figure \ref{CCF2} shows the result concerning the S/N and, again, it shows no significant detection.

%
%-------------------------------------------------------------
%                 A Fig.as large as the width of the column
%-------------------------------------------------------------

\begin{figure}
   \centering
   \includegraphics[width=\hsize]{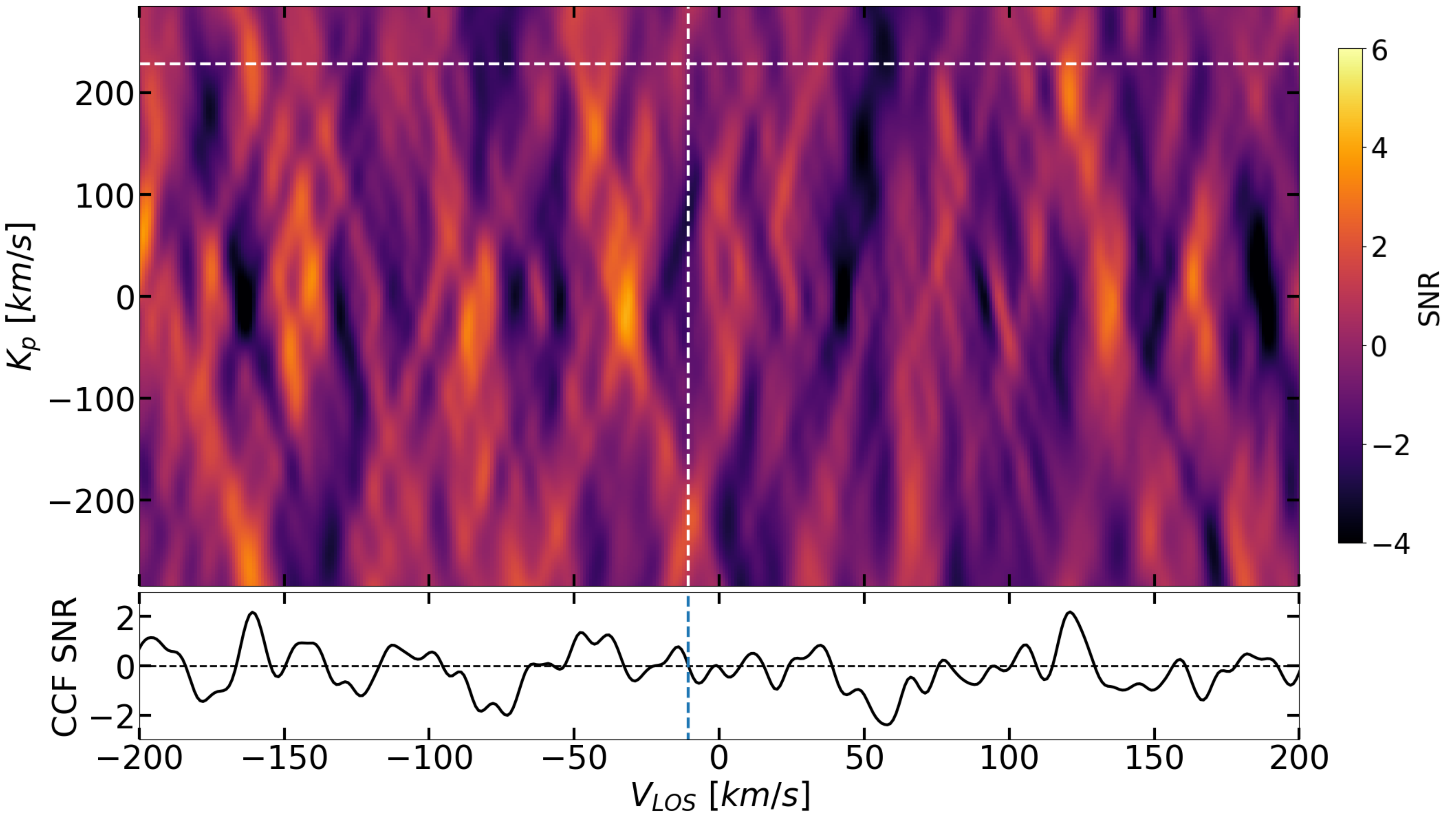}
      \caption{CCF constructed using templates for Na, K, FeH, V, and TiO. The x-axis indicates the offset relative to the line of sight velocity of the stellar system ($V_{LOS}$), while the y-axis displays the range of Doppler velocities ($K_{p}$) associated with the planet's orbit around the star. The CCFs are presented in S/N units to facilitate interpretation. White lines mark the predicted velocities for both the star and the planet, serving as a guide. The bottom part of the figure zooms in on the CCF at the expected $K_{p}$ velocity, offering a detailed view of the search for atmospheric signals.}
         \label{CCF3}
   \end{figure}

In our analysis, we produced a plot \ref{CCF3} similar to those commonly seen in high-resolution transmission spectroscopy (HR TS) studies, yet with varied assumptions regarding the Doppler shift of the planet, denoted as $K_p$.
These maps involve identifying the planet's velocity signal across a range of potential orbital velocities ($K_p$) and systemic velocities ($V_{sys}$) by collapsing the CCF into a two-dimensional (2D) space.
This variation is instrumental in probing different potential parameters of the exoplanetary system and identifying any systematic errors or stellar variability signals. Figure \ref{CCF3} illustrates these results, presented in terms of the S/N. The white dotted line in this plot serves as a marker for the expected positions in the $K_p$-$V_{sys}$ plane, typically used in HR TS to correlate the planetary signal with its systemic velocity. Notably, our analysis, as visualised in this plot, does not reveal any discernible signal. For an in-depth understanding of the transition from CCF maps to $K_p$-$V_{sys}$ maps, we refer to HR TS papers such as \citet{2017A&A...602A..36W, 2018A&A...616A.151C}, which delineate the methodology and interpretation in greater detail.

%-------------------------------------------------------------
%                 A Fig.as large as the width of the column
%-------------------------------------------------------------

\section{Analysis}

To place our results into context, we performed an injection and recovery test in order to assess the quality and limitations of the data. The predictions were made using data sets drawn from frames outside the transit. First, we generated a model with petitRADTRANS, which is convolved by a Gaussian kernel accounting for the rigid-body rotation of the planet on its axis, the equivalent ESPRESSO signal convolution, and the smearing of the absorption lines due to the change in the line-of-sight radial velocity during each given exposure. Second, we combined all the spectra from the out-of-transit data to create a master spectrum as before, with which we injected the predicted signal generated previously. Then we inject photon noise to get the equivalent S/N expected from an exposure.
The synthetic spectrum with the noise is defined as follows: 

\begin{align}
 f_{i, \rm{syn}}(\lambda) = f_{\textrm{m}_i}(\lambda) - \frac{r^2_{planet}(\lambda)}{R_{star}^2} + A \frac{\epsilon_{i}(\lambda)}{\sqrt{f_{\textrm{m}_i}(\lambda)}},
\end{align}

\noindent where $r$ is the planet size predicted by petitRADTRANS, $\epsilon_{i}$ is a random number drawn from a normal distribution $\mathcal{N}(0,1)$, and $A$ is a free parameter that scales the model noise to that of the real continuum level, as measured in the combined data. This was done to simulate as much as possible the noise of a real frame with the exposure time taken.

%-------------------------------------------------------------
%                 A Fig.as large as the width of the page
%-------------------------------------------------------------
   \begin{figure*} 
   \centering
   \includegraphics[width=17cm]{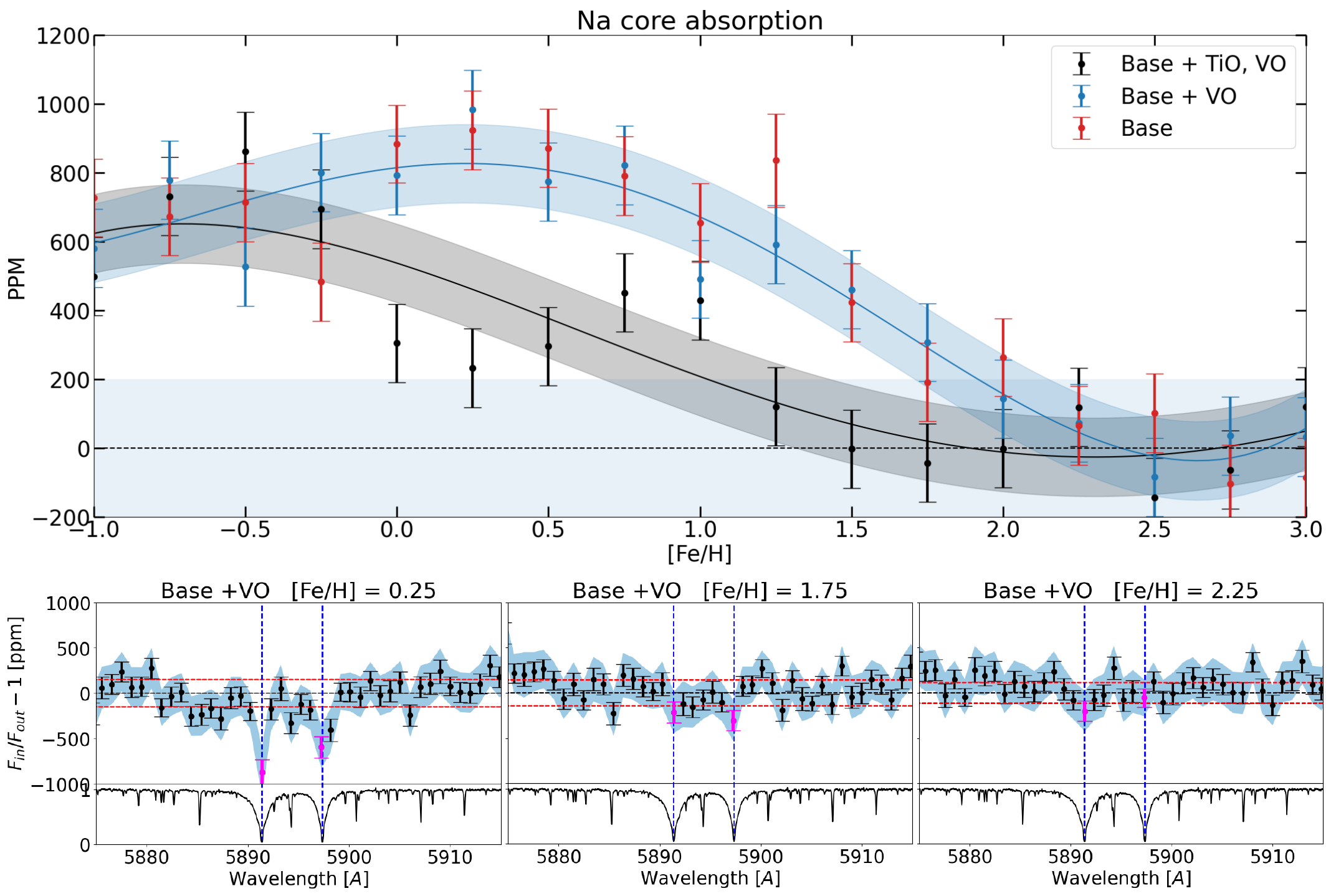}
      \caption{Top: Transmission Spectroscopy showing extra absorption in the Na doublet's core lines across different global abundances, measured in PPM. This plot compares three injected models, illustrating the presence or absence of TiO and VO. It features one example with error bars plus a main strand after several runs. In light blue, regions below 200 ppm are highlighted, indicating areas under the estimated error threshold coming from models without signals. Bottom: Three examples of Transmission Spectroscopy in PPM are presented. Black points represent data binned by a factor of 100, the light blue area denotes a 95\% confidence interval, magenta highlights the core lines, and red dashed lines indicate the standard deviation of the continuum. Below these plots are normalised flux profiles centred on the core lines, with a dashed blue line marking the Na doublets}
         \label{Transmission_Syntetic_1}
   \end{figure*}
%

%-------------------------------------------------------------
%                 A Fig.as large as the width of the page
%-------------------------------------------------------------
   \begin{figure*} 
   \centering
   \includegraphics[width=17cm]{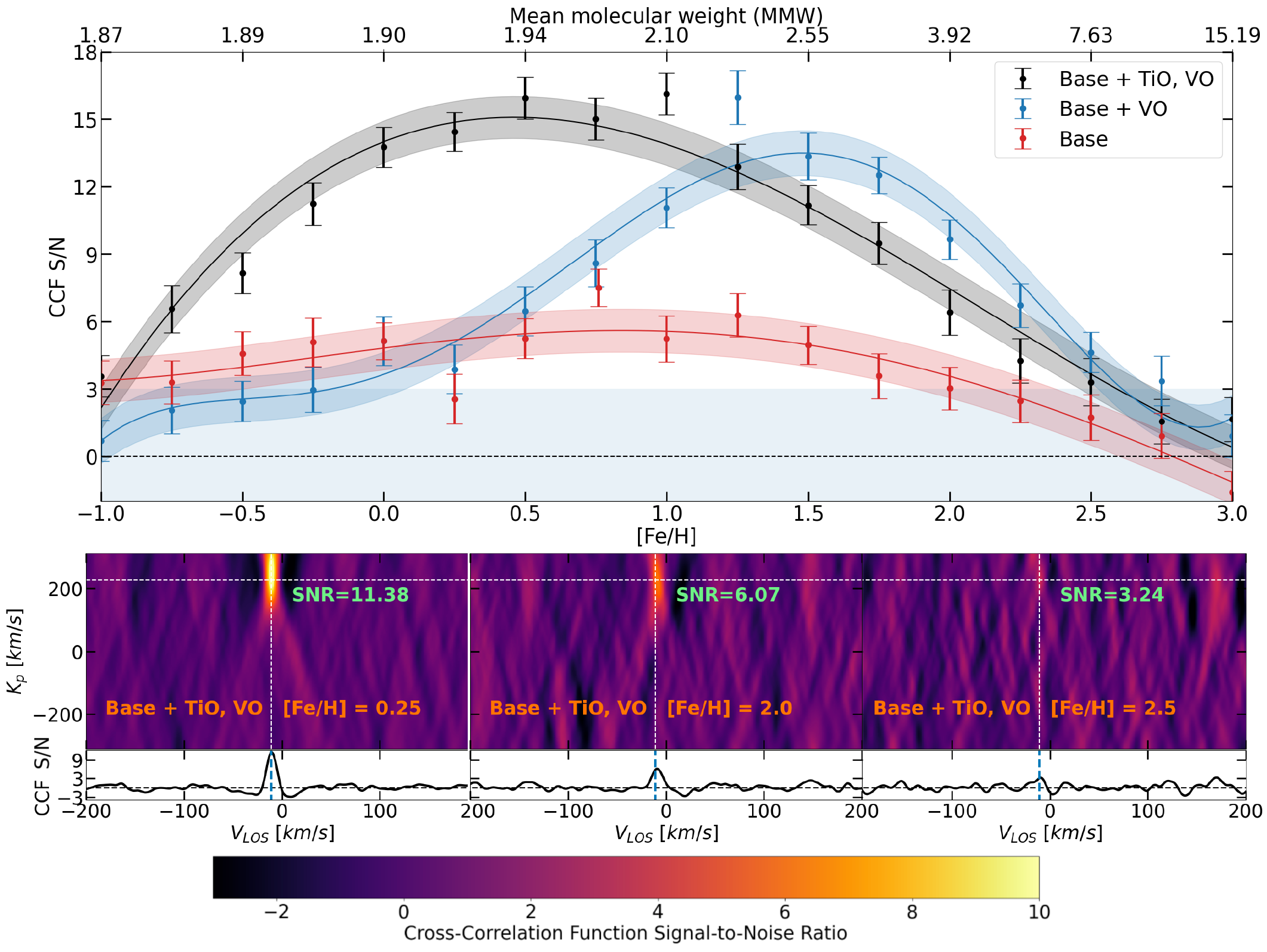}
      \caption{
      Comparison of the cross-correlation signal-to-noise ratio (S/N) for atmospheres with varying compositions and metallicities.
      Top: CCF in terms of S/N for different abundances or the equivalent Mean molecular weight. It compares three different injected models with the presence or absence of TiO and VO. The bottom plots give CCF examples for comparison of the planet's Keplerian velocity vs the line-of-sight velocity map. It shows different levels of significance for different abundances. Below them is a plot of a row in the injected planet velocity from the CCF velocity map}
         \label{Transmission_Syntetic_2}
   \end{figure*}

It is important to note that in our injection-recovery tests, the simulated planetary signals were injected into the master out-of-transit spectrum, which combines the observed spectra after telluric correction but before the removal of instrumental systematics, such as the 'wiggles' discussed earlier in this work. This approach allowed us to work with a clean spectrum that minimises the influence of systematics and provides a baseline for the ideal detectability of atmospheric features.
However, this method does not fully account for all potential residual effects from telluric absorption and instrumental systematics, particularly concerning species retrieval, especially for water. Water vapour, the primary contributor to telluric absorption in our wavelength range, is not included in our cross-correlation templates, resulting in minimal impact on species detection. Therefore, the main limitations to our detection sensitivity stem from photon noise and any residual systematics that persist after correction.

Figure \ref{Transmission_Syntetic_1} presents the outcomes from various synthetic data simulations around the sodium doublet, utilising three distinct petitRADTRANS models. The 'base' model incorporates elements as shown in Fig. \ref{Fig_MassFraction2}, albeit with the abundances of TiO and VO set to zero. This base model also stipulates a C/O ratio of 0.5 and an equilibrium temperature of 2000K at 1 mbar. The other two models, in contrast, include either TiO or VO and TiO together.

In the top panel of Fig. \ref{Transmission_Syntetic_1}, we display the mean line core parts per million (PPM), highlighted in magenta, across varying metallicities ([Fe/H]). The bottom panel, drawing parallels with Fig. \ref{Fig_Sodium_NoDetection}, showcases three examples of transmission spectroscopy. The error bars for each model are derived from the dispersion in the continuum, which is denoted by the red dotted lines in these examples.

A noteworthy observation from our analysis is the differing detectability trends between the models. While the Base and Base+VO models exhibit similar detectability patterns, the Base+TiO and VO combination diverges, showing reduced detectability at a lower metallicity threshold of 1.25. To quantitatively assess these trends, a polynomial fit was applied to each model, represented by solid lines in the figures. The area around these fitted lines, equivalent to the thickness of the error bars, was measured to gauge the uncertainties in our model predictions. We have marked the 200 ppm threshold as a reference point for non-detectability, based on observations that several consecutive runs exhibited fluctuations exceeding this value in the line core dispersion. This comprehensive approach, combining polynomial fitting with error analysis, enhances the reliability of our detectability assessments.

Given the absence of a discernible signal in the sodium doublet, as shown in Fig. \ref{Fig_Sodium_NoDetection}, we proceeded to estimate an upper limit for the metallicity of the planetary atmosphere. For the scenario hypothesizing an atmosphere devoid of clouds, our calculations suggest an upper metallicity limit of 1.0 in the case of an atmosphere containing VO. For other models under consideration, this upper limit is estimated to be around 2.0. These limits are contingent on the non-detection of the sodium doublet and aim to provide a constraint on the atmospheric composition.

Figure \ref{Transmission_Syntetic_2} presents a comprehensive comparison of the CCFs for the three distinct models at the top of the figure. This comparison not only highlights the S/N detected at varying metallicities but also aligns this with the equivalent mean molecular weight (MMW) for each model: Base, Base + TiO, and Base + Tio + VO. The S/N is calculated based on the standard deviation (STD) observed in the CCF map's regions, excluding the areas with the injected signal.

The lower part of Fig. \ref{Transmission_Syntetic_2} illustrates three representative examples, aligning with the pattern established in Fig. \ref{CCF3}. These examples are further elucidated by a polynomial fit, which delineates the primary trend and is represented by a line with a thickness corresponding to the error bars. Additionally, areas exhibiting an S/N of less than 3 are marked in light blue, signifying the threshold for detectability. This dual analysis of CCFs and S/N across different models and metallicities offers a nuanced understanding of our detectability limits.

It is worth mentioning that in the case of the base+VO models for the cross-correlation, we observe compatibility with the data at both high metallicities and metallicities below [Fe/H] = 0.25. At low metallicities, the decreased abundance of VO leads to fewer detectable spectral lines, reducing the cross-correlation signal below our detection threshold. However, we consider this low-metallicity scenario to be unlikely, as it would require the planet's atmosphere to have a metallicity lower than that of its host star ([Fe/H] = 0.12 $\pm$ 0.10 dex). Combined with our non-detection of Na, which would be more detectable at lower metallicities with a larger scale height, we focus our interpretation on the high-metallicity regime, which is more consistent with both our data and theoretical expectations.

%Adopting a methodology akin to that used in transmission spectroscopy, 
We established a lower metallicity limit for the CCF analysis. For all the models examined, this limit is determined to be approximately 2.25. The increase primarily influences this threshold in the mean molecular weight (MMW), which leads to a flattening of the spectrum.

\section{Discussion}

The transmission spectrum of the sodium doublet, anticipated to be the strongest signal based on models, lacks sufficient S/N  to confirm its detection. However, this does not rule out the presence of sodium in these atmospheric layers, as various factors (e.g. cloud cover or a high metallicity atmosphere that lowers the atmospheric scale height) can diminish the signal and flatten the spectrum. On the other hand, the non-detection of any H${\alpha}$ signal suggests there is no dense hydrogen cloud around the planet due to photoevaporation. 

The absence of a detectable signal in the CCF suggests an atmosphere-dependent upper metallicity limit, assuming no cloud cover. In analyzing three different atmospheric models, each presenting distinct structures in transmission spectroscopy considering the presence or absence of TiO and VO, we observed notable differences. Specifically, an atmosphere devoid of both TiO and VO manifests a significantly reduced number of spectral lines, thereby diminishing the CCF's detectability. This effect is exemplified by the relatively constant red line in Fig. \ref{Transmission_Syntetic_2}, up until the point where an increase in MMW and a corresponding decrease in scale height lead to a flattening of the spectral lines.

Conversely, introducing VO into the atmosphere enhances detectability, peaking at a metallicity of 1.5, before experiencing a decline due to similar flattening effects. The inclusion of TiO, known for its relatively strong spectral lines in the optical, creates a detectability peak around a metallicity of 0.0. However, this detectability subsequently diminishes at higher MMW values. These findings underscore the significant impact of atmospheric composition on the CCF's detectability in exoplanetary studies.
Our analysis suggests two plausible explanations for the observed results, as explained below.

Firstly, the non-detection of H$\alpha$, which could extend to the Roche-lobe, indicates no evidence of overflow. This suggests that the atmosphere is not undergoing significant Roche-lobe overflow despite the planet's position within the Neptune Desert. However, it is still possible that the mean density of the atmosphere is higher than initially anticipated, leading to a significantly reduced scale height (H) \citep{2013ApJ...775...80F}. Such a dense atmosphere could fall below our CCF method's detection threshold. This scenario might indicate a planet subjected to substantial stellar wind.

Secondly, the presence of clouds or hazes in the upper atmosphere could be obscuring the signatures from deeper atmospheric layers ($P \geq 10^{-4} $bar). This obstruction by clouds would limit the technique's ability to probe below these cloud regions, a phenomenon observed in other exoplanetary studies such as those detailed in \citet{2014Natur.505...69K} and \citet{2014Natur.505...66K}. Supporting evidence for clouds in the planet's atmosphere comes from CO detections made using Spitzer
%\citep{2020ApJ...903L...6D}
and James Webb Space Telescope
(JWST; \citealt{2020ApJ...903L...6D, 2024ApJ...962L..20R}). Furthermore, recent optical data from CHEOPS \citep{2021ExA....51..109B} report a high geometric albedo of approximately 0.8, indicative of a high-altitude metallic cloud layer \citet{2023A&A...675A..81H}).

Our data do not provide conclusive evidence to definitively support either explanation. On one side, Roche-lobe overflow models suggest significant atmospheric escape under a basic stellar wind model \citep{2017ApJ...835..145J}. Additionally, the low orbital eccentricity implies a lack of evidence for recent migration, suggesting the planet has likely been exposed to stellar wind for over 1 Gyr. This prolonged exposure could lead to the loss of light elements from the upper atmosphere, especially if the star has been active for a significant duration. This scenario does not even account for potential influences from Coronal Mass Ejections (CMEs) or strong flaring events, which could further contribute to atmospheric stripping.

Conversely, observations from CHEOPS indicating high-altitude clouds suggest an alternative mechanism at play \citep{2021ExA....51..109B}. It is crucial to note that photoevaporation, a key process in atmospheric loss, is most vigorous in the first 100 Myr of a system's life when the star is highly active. Therefore, the current rate of evaporative mass loss may be relatively low, but historical events, including active outbursts, CMEs, or strong flaring, could have significantly influenced the atmosphere's composition and structure (photochemistry and hydrogen escape). Given these considerations, both explanations remain viable, and additional data will be key to discerning between these two scenarios.

The recent JWST observations by \citet{2024ApJ...962L..20R} provide constraints on planetary metallicity within the framework established by \citet{2019ApJ...874L..31T}, setting an upper limit of \( Z_p = 90\% \pm 2\% \)—approximately 850 times solar metallicity, or \([\text{Fe/H}] < 2.93\). In contrast, \citet{2013ApJ...775...80F}, as referenced in \citet{2024ApJ...962L..20R}, proposed a lower limit of \( Z_p = 21\% \) (around 20 times solar), based on the absence of planets below this threshold in population synthesis predictions. These JWST NIRISS data reveal muted spectral features, consistent with a water- and methane-dominated atmosphere, which are well-aligned with these metallicity limits.

These limits align well with our inferred lower limit of \([\text{Fe/H}] \geq 2.25\) (or \(\geq 180 \times\) solar). Both studies indicate the presence of high-altitude clouds or hazes that could obscure deeper atmospheric layers, consistent with the high geometric albedo reported by \citet{2023A&A...675A..81H}. The lack of significant molecular features in both our optical observations and their near-infrared data supports the hypothesis of a metal-rich atmosphere with clouds at millibar pressures. Moreover, the absence of significant atmospheric escape signatures in both studies suggests that \planet\ retains its atmosphere despite extreme irradiation, possibly due to the weak high-energy emission from its host star. Therefore, our ground-based high-resolution observations complement the space-based JWST observations, collectively enhancing our understanding of this ultra-hot Neptune.

Our study further refines the lower limits, depending on the presence of TiO and VO in the atmosphere. If TiO and VO are detected, the upper limit is \([\text{Fe/H}] > 2.5 \pm 0.25\); without these elements, the limit is \([\text{Fe/H}] > 2.25 \pm 0.25\). \citet{2024ApJ...962L..20R} report no evidence of TiO and VO, suggesting that the latter scenario is more plausible. Nonetheless, these limits are approximate, based on models that assume solar-like elemental ratios for metals. A joint fit using models tailored to individual molecules would enable a more precise characterisation.

\section{Conclusions}

\planet, classified as an ultra-hot Neptune and situated deep within the Neptune desert, presents a unique opportunity for atmospheric study due to its high temperature, magnitude, and confirmed atmosphere. Utilising ESPRESSO's exceptional sensitivity, resolution, and photon-collecting capability, we aimed to acquire spectral measurements of the planet's atmosphere through transmission spectroscopy. Despite addressing numerous systematics in the ESPRESSO data and analysing three distinct transit events, we did not find any statistically significant evidence of an atmospheric signal in the sodium doublet. Instead, we established lower metallicity limits of [Fe/H] = 2.25 for all models, reflecting the complexities of detecting atmospheric signatures in such environments.

Our search for H-alpha signals, indicative of potential atmospheric winds, is consistent with the lack of strong X-ray emission from the star that could lead to atmospheric mass loss \citep{2024MNRAS.527..911F}. The non-detection of atmospheric signatures using the cross-correlation technique, even when combining multiple expected elements, aligns with a possible high-metallicity atmosphere scenario. This would imply a high mean molecular weight, subsequently reducing the atmospheric scale height below our current detection threshold. Alternatively, the presence of a cloud or haze layer in the upper atmosphere, obscuring deeper layers rich in strong opacities, remains a viable explanation.

These findings are in agreement with the recent JWST observations by \citet{2024ApJ...962L..20R}, who also reported muted spectral features in the transmission spectrum of \planet. Their analysis suggests atmospheric metallicities between 20 and 850 times solar, similar to our inferred lower limit. Both studies indicate the presence of high-altitude clouds or hazes that could obscure deeper atmospheric layers, consistent with the high geometric albedo reported by \citet{2023A&A...675A..81H}. The lack of significant molecular features in both our optical observations and their near-infrared data supports the hypothesis of a metal-rich atmosphere with clouds at millibar levels of pressure.

The consistency between our ESPRESSO observations and the JWST data underscores the importance of combining ground-based and space-based observations for a comprehensive understanding of exoplanet atmospheres. Future observations, including those with JWST NIRSpec and further high-resolution spectroscopy, will be crucial to further constrain the atmospheric properties of \planet\ and explore the mechanisms behind atmospheric retention in ultra-hot Neptunes.

\begin{acknowledgements}
JSJ gratefully acknowledges support by FONDECYT grant 1240738 and from the ANID BASAL projects ACE210002 and FB210003.
We sadly note the passing of our colleague and co-author Yakiv Pavlenko (YP), whose contributions to this work and the field of astrophysics remain foundational. His investigations were carried out under the MSCA4Ukraine programme (project number 1.4-UKR-1233448-MSCA4Ukraine), funded by the European Commission. His dedication and insight continue to inspire those who had the privilege of collaborating with him.
\end{acknowledgements}

\bibliographystyle{aa}
%\bibliography{LTT9779_2021a}
\bibliography{aanda}

\end{document}